\documentstyle[prb,aps,eqsecnum,twocolumn]{revtex}
\def\break#1{\pagebreak \vspace*{#1}}
\begin{document}
\draft
\title{Charging effects in quantum wires}
\author{Reinhold Egger and Hermann Grabert}
\address{Fakult\"at f\"ur Physik, Albert-Ludwigs-Universit\"at,
Hermann-Herder-Stra{\ss}e 3, D-79104 Freiburg, Germany}
\date{to be published in Phys. Rev. B}
\maketitle
\widetext
\begin{abstract}
We investigate the role of charging effects in a voltage-biased
quantum wire.  Both the finite range of the
Coulomb interaction and the long-ranged nature of the
Friedel oscillation imply a finite 
capacitance, leading to a charging energy. 
While observable Coulomb blockade effects
are absent for a single impurity, they are
crucial if islands are present. For a double barrier,
we give the resonance condition, fully taking into account the
charging of the island.
\end{abstract}

\narrowtext

\section{Introduction}

One-dimensional (1D) quantum wires have attracted 
much interest lately due to theoretical developments and
advanced fabrication techniques.  By using a
special split-gate technology \cite{tarucha95} or
cleaved-edge overgrowth,\cite{yacoby96} high-quality 1D channels in GaAs-AlGaAs
heterostructures have recently been fabricated. The  typical mean free
path can be of the order of $10 \mu$m, which brings one close to
the possibility of experimentally probing 
the transport properties of quantum wires,
with either none or only very few impurities. The most prominent
transport quantity, the conductance, has been theoretically 
studied in great detail, especially for the clean case \cite{oned,egger96}
or for a single impurity.\cite{egger96,kane92,matveev93,fendley95}
However, little effort has been undertaken so far to reveal
the nature of charging effects \cite{sct} in a quantum wire.
In this paper, we discuss charging effects
for the simplest case of a spinless single-channel wire.

The importance of charging effects is established by the 
magnitude of the charging energy $e^2/C$. To calculate this
quantity, we utilize standard
bosonization methods \cite{peschel75,emery79,haldane81} in conjunction 
with our recently developed boundary condition approach.\cite{egger96} 
This formalism allows for
a particularly simple derivation of the capacitance $C$ of a single impurity
in the limit of  strong impurity backscattering strength. 
It is intuitively clear that the capacitance will to a large degree 
depend on the interaction range $R$
of the Coulomb potential in the quantum wire.\cite{maura}
Therefore, to describe
charging effects for realistic experimental setups characterized by a
 finite $R$, we go beyond the strict Luttinger liquid 
picture\cite{haldane81} which has
zero range, $R=0$, and consider an
arbitrary screened Coulomb interaction potential $U(x-y)$. It turns out that
(at least concerning dc properties) 
charging effects will only be present if the quantum wire contains
islands formed by two impurities. Including the charging contribution, we
 provide the resonance condition for resonant
tunneling through a double barrier structure. 
 Due to the finite charging energy, 
we find a different resonance condition 
than the one predicted\cite{kane92,furusaki93}
for resonant tunneling in a Luttinger liquid.
 
\break{1.0in}

The outline of this paper is as follows. In Sec.~\ref{sec:2},
the boundary condition approach is applied to the case of a
quantum wire with arbitrary Coulomb interactions containing
a single impurity.   This formalism is used in Sec.~\ref{sec:3}
to compute the charge screening cloud around a strong impurity, from
which one can define the capacitance $C$.  The resonance condition
for resonant tunneling through a double barrier is derived in 
Sec.~\ref{sec:4}, and the additional charging contribution is 
discussed in detail. Finally, some conclusions are offered in
Sec.~\ref{sec:5}.

\section{Boundary condition formalism} \label{sec:2}

We employ the standard bosonization method,\cite{peschel75,emery79,haldane81}
which holds in the low-energy regime where only plasmons are
well-defined eigenmodes. The electron creation operator
can then be written in terms of plasmon displacement
 fields $\theta(x)$ and $\phi(x)$,
\begin{equation}\label{bo}
\psi^\dagger(x) =  \sqrt{\frac{\omega_c}{2\pi v_F}} \sum_{p=\pm}
\exp\{ ipk_F x + i\sqrt{\pi} [p\theta(x)+\phi(x)]\} \;,
\end{equation}
where $\omega_c=v_F k_F$ is the bandwidth (we put $\hbar=1$),
and the sum goes over left- and right-moving pieces ($p=\pm$).
These fields obey the algebra
\[
[ \phi(x), \theta(y) ] = -(i/2) {\rm sgn}(x-y) \;,
\]
such that $\Pi=\partial_x \phi$  constitutes the 
canonical momentum for the $\theta$ field. The 
non-interacting 1D electron gas is then described by the
bosonized form of the massless Dirac Hamiltonian
\begin{equation}\label{h0}
H_0 = \frac{v_F}{2} \int dx \left [ 
\Pi^2(x) + (\partial_x \theta(x))^2 \right] \;.
\end{equation}

Arbitrary screened Coulomb interactions can be included as follows.
From Eq.~(\ref{bo}), the  boson representation of 
the density operator is 
 \begin{equation}\label{dens}
\rho(x) = \frac{k_F}{\pi} + \frac{1}{\sqrt{\pi}} \partial_x \theta(x)
+ \frac{k_F}{\pi} \cos[2k_F x+ 2\sqrt{\pi} \theta(x)]\;.
\end{equation}
The $2k_F$-oscillatory component comes from interference of 
right- and left-movers and is responsible for a
Friedel oscillation \cite{egger95} in a system with broken
translational invariance.
In the standard expression for the electron-electron interaction,
\begin{equation}\label{int}
H_I = \frac12 \int dx dy \,\rho(x) U(x-y) \rho(y) \;,
\end{equation}
the  interaction among the $2k_F$ parts of
$\rho(x)$ and $\rho(y)$ gives rise to electron-electron
backscattering.\cite{emery79}
 In the following, we assume that the screened
interaction  
\[
U(x-y) = \int \frac{dk}{2\pi}\, U_k \exp[ik(x-y)]
\]
is sufficiently long-ranged such that
 $U_{2k_F}$ is small. In that case, backscattering can be
neglected. This is of course not an essential assumption but simplifies
notation in the following.
In a homogeneous wire, the interaction between the slow and
the $2k_F$ part of $\rho$ averages out due to a rapidly
oscillating phase factor, i.e., due to momentum  conservation.  We then
take into account only the forward scattering contribution,
\[
H_I  = \frac{1}{2\pi} \int dx dy \,
\partial_x \theta(x) \,U(x-y) \,\partial_y \theta(y)\;.
\]
This is the interaction  between
the slow components in Eq.~(\ref{dens}).

Next we discuss how to incorporate coupling of the quantum wire
to external reservoirs.  Incidentally, it has become quite customary
to use non-interacting 1D Luttinger liquids described by Eq.~(\ref{h0})
as a model for the external leads.\cite{oned,oned2}
There is a serious shortcoming
inherent to such a modelling. If one connects
the interacting wire to 1D non-interacting leads, one effectively has
an inhomogeneous interaction potential $U(x,y)$. 
In that case, the Coulomb interaction between the slow and the $2k_F$ parts
of $\rho$ does {\em not}\, necessarily average to zero since 
momentum is not conserved anymore.  In fact,
if the interactions are switched off on a lengthscale $1/k_F$, we
obtain from Eq.~(\ref{int}) effective potential scatterers 
at the boundaries of the wire of large backscattering strength,  
\[
V_{\rm eff} \simeq \left(\frac{1}{g^2}- 1\right) \omega_c \;.
\]
Hence, from a microscopic point of view,
transport would practically be suppressed with such leads.
This additional interaction term has been disregarded in 
Refs.\onlinecite{oned,oned2}, but it is present in a quantum wire
connected to 1D non-interacting leads.  

These difficulties are avoided by the boundary condition 
approach of Ref.\onlinecite{egger96}. The reservoirs inject currents at both
ends of the wire. In the spirit of Landauer's approach for
non-interacting systems,\cite{landauer57} the injection of currents
can be described by Sommerfeld-type radiation conditions.
Coupling to external reservoirs held
at a chemical potential difference $\Delta\mu=eU_{\rm ex}$  leads
to the boundary conditions 
\begin{equation} \label{bc0}
\left ( \pm \frac{\partial}{\partial x} + 
\frac{1}{v_F}\frac{\partial} {\partial t} \right)
\langle \theta(x\to \mp \infty,t)\rangle = eU_{\rm ex}/2\sqrt{\pi}v_F\;.
\end{equation}
These conditions have to be imposed 
at both ends of the quantum wire, i.e.,
far away from all impurities but still inside the quantum wire such that 
bosonization is meaningful, and for all times $t$. 
In general, the effects of an applied voltage cannot be captured by adding
new terms to the Hamiltonian.
 The boundary condition approach holds for
arbitrary Coulomb interaction
potentials and allows for arbitrary arrangements of impurities inside
the quantum wire. Since we are not directly concerned with conductance
calculations in the following, we use for simplicity
the $T=0$ imaginary-time formalism, where the equivalent conditions 
\begin{equation} \label{bc}
\left ( \pm \frac{\partial}{\partial x} + 
\frac{i}{v_F}\frac{\partial} {\partial \tau} \right)
\langle \theta(x\to \mp \infty,\tau=0)\rangle = eU_{\rm ex}/2\sqrt{\pi}v_F
\end{equation}
are imposed. Here we have picked the time $\tau=0$ by convention.

We first discuss the case of a single impurity located at $x=0$, and consider
the generating functional 
\[
Z(x_0,s)=\langle
 \exp[2\sqrt{\pi} i s \theta(x_0,\tau=0)]\rangle\;.
\]
One can formally solve for $Z$ by using the auxiliary field 
$q(\tau)=2\sqrt{\pi} \theta(0,\tau)$, where the constraint is enforced by
a Lagrange multiplier field $\Lambda(\tau)$. The
resulting effective action is
\begin{eqnarray*}
S_e[\theta,\Lambda,q] &=& \frac{v_F}{2} \int d\tau dx  \left[
\frac{1}{v_F^2} (\partial_\tau \theta)^2 + (\partial_x \theta)^2 \right]  \\
&+& \frac{1}{2\pi} \int d\tau dx dy \,\partial_x \theta(x)
U(x-y) \partial_y\theta(y) \\
&+& V \int d\tau \cos [q(\tau)] 
- 2\sqrt{\pi} i s\theta(x_0,0) \\&+&
i\int d\tau \Lambda(\tau) [2\sqrt{\pi} \theta(0,\tau)-q(\tau)] \;,
\end{eqnarray*}
where $V$ denotes the impurity backscattering strength.
The action is now quadratic in the $\theta$ part, which therefore  can be 
integrated out by solving the classical
Euler-Lagrange equation. This has to be done 
under the boundary condition (\ref{bc}).

Analogous to the zero-range Luttinger liquid case 
treated in Ref.\onlinecite{egger96},
this can be achieved by decomposing fields into homogeneous and
particular parts, $\theta=\theta_h+\theta_p$ and
$\Lambda=\Lambda_h + \Lambda_p$. The particular solution 
$\theta_p$ has to fulfill Eq.~(\ref{bc}) and the Euler-Lagrange equation
\begin{eqnarray}\label{euler}
&& \frac{1}{v_F^2} \partial_\tau^2 \theta_p(x,\tau) +
\partial_x^2 \theta_p(x,\tau)   \\ \nonumber \quad &+&
 \int dy \frac{U(x-y)}{\pi v_F}
\partial_y^2 \theta_p(y,\tau) = (2\sqrt{\pi}i/v_F)
\Lambda_p (\tau) \delta(x) \;. 
\end{eqnarray}
A solution subject to Eq.~(\ref{bc}) requires a 
$\tau$-independent $\Lambda_p$. We make the ansatz
\begin{equation}\label{ansatz}
\theta_p(x,\tau) = - \frac{e\bar{\varphi}}{2\sqrt{\pi}v_F} |x| -
i\tau \frac{e(U_{\rm ex}-\bar{\varphi})}{2\sqrt{\pi}} + f(x) - f(0) \;,
\end{equation}
where $\partial_x f(x) \to 0$ as $|x|\to \infty$ for the 
$\tau$-independent function $f(x)$. 

 Switching to Fourier space,  one finds from Eqs.~(\ref{euler})
and (\ref{ansatz}) 
\begin{eqnarray}\label{particular}
-\frac{e\bar{\varphi}}{\sqrt{\pi} v_F} ( 1+ U_k/\pi v_F ) 
 &-& k^2  ( 1+ U_k/\pi v_F ) f_k 
\\ &=& (2\sqrt{\pi}i/v_F ) \,\Lambda_p \;. \nonumber
\end{eqnarray}
The $k=0$ component of Eq.~(\ref{particular}) determines the
quantity  $\bar{\varphi}$ in terms of the zero mode $\Lambda_p$ of
the Lagrange multiplier,
\[
\Lambda_p = \frac{ie\bar{\varphi}}{2\pi g^2} \;.
\]
Here we have introduced the usual dimensionless Luttinger liquid
parameter $g$ as a measure of the
forward scattering interaction
strength,\cite{kane92,peschel75,emery79,haldane81}
\[
1/g^2 = 1+ U_0/\pi v_F \;.
\]
The non-interacting limit is $g=1$, and for repulsive interactions, 
one has $g<1$.
Naturally, a screened interaction is characterized by $g$ and the 
range $R$ (and possibly by other parameters). For $R=0$, the
$k\neq 0$ components of Eq.~(\ref{particular}) vanish identically,
and $f(x)$ stays constant. In the general finite-range case, we have
from Eq.~(\ref{particular}) the simple result 
\begin{equation}\label{fk}
f_k =  \frac{e\bar{\varphi}(U_0-U_k)}{\pi^{3/2} \omega_k^2}\;,
\end{equation}
which apparently vanishes for the zero-range case where $U_k=U_0$ for all $k$.
Here, the plasmon frequency is 
\begin{equation}\label{wk}
\omega_k=v_F|k|\sqrt{1+U_k/\pi v_F}\;.
\end{equation}

The homogeneous solution (for $U_{\rm ex}=0$ boundary condition)
 is easily expressed in terms of the  boson propagators
\begin{equation}\label{bos}
F(x,\omega) = v_F \int dk \frac{\exp(ikx)}{\omega^2 + \omega_k^2}\;,
\end{equation}
 such that
\[
\theta_h(x,\tau) = -\frac{i}{\sqrt{\pi}} \int \frac{d\omega}{2\pi}
e^{i\omega\tau} [\Lambda_h(\omega)F(x,\omega) - s F(x-x_0,\omega)]\;.
\]
Inserting $\theta_h+ \theta_p$ back into the action, one observes that
$\Lambda_h$ appears only in quadratic form and can therefore
be integrated out immediately. After some algebra,
 one finally arrives at 
\begin{equation}\label{gen}
 Z(x,s) =W(x)^{s^2} \left\langle e^{2\sqrt{\pi}i s\theta_p(x,0)} 
\exp\left[ is \int \frac{d\omega}{2\pi} q(\omega) 
\frac{F(x,\omega)}{F(0,\omega)} \right] \right\rangle\;.
\end{equation}
The envelope function $W(x)$ is given by 
\begin{equation}\label{envelope}
W(x) = \exp \left[ \int \frac{d\omega}{2\pi}\frac{F^2(x,\omega)-
F^2(0,\omega)}{F(0,\omega)} \right] \;,
\end{equation}
and the average over $q(\omega)=(2\pi)^{-1} \int d\omega \,q(\tau) 
\exp(i\omega \tau)$ has to be carried out using the action
\begin{eqnarray*}
S  & =&  \int \frac{d\omega}{2\pi} \frac{q(\omega) q(-\omega)}{4F(0,\omega)}
+ \frac{e\bar{\varphi}}{2\pi g^2} \int d\tau \,q(\tau)\\
 &+& V \int d\tau \cos[q(\tau)-ie(U_{\rm ex}-\bar{\varphi})\tau] \;.
\end{eqnarray*}
In principle, $\bar{\varphi}$ is a fluctuating quantity: One has to
average over it, because it is the zero-mode of the Lagrange multiplier field.
Its physical meaning is the four-terminal voltage.\cite{egger96}
In the case of strong impurities considered here, $\bar{\varphi}$
is therefore just equal to the two-terminal voltage $U_{\rm ex}$.
In the following, we restrict ourselves to the strong-impurity limit 
where charging effects are most pronounced, and put 
$\bar{\varphi}=U_{\rm ex}$.

\section{Impurity screening profile and capacitance} \label{sec:3}

Let us now analyze the expectation value of the 
slow component   of the density, $\rho_0(x)=\langle 
\partial_x\theta\rangle/\sqrt{\pi}$.
 It can be obtained by suitable differentiation of the generating
functional (\ref{gen}), with the
result $\rho_0(x) = \partial_x \theta_p(x,0) /\sqrt{\pi}$,
since the contributions from the homogeneous solution cancel out.\cite{foot1}
Therefore, Eq.~(\ref{ansatz}) yields
\begin{equation}\label{slow}
\rho_0(x)= - \frac{eU_{\rm ex}}{2\pi v_F} \,{\rm sgn}(x) +
\frac{1}{\sqrt{\pi}} \, \partial_x f(x) \;.
\end{equation}
The Fourier transform of $f(x)$ is given in Eq.~(\ref{fk}).
The uniform contribution $\sim {\rm sgn}(x)$ is due to charging of the large
shunt capacitances between the quantum wire and the metallic
(screening and confining) gates, and the corresponding charge on the impurity, 
$Q_s=Le^2 U_{\rm ex}/4\pi v_F$, scales with the total length $L$ of 
the wire. The observation of 
dc charging effects for a single impurity is rendered impossible
by this macroscopically large charge.\cite{egger96} This can also be
shown by computing the current-voltage characteristics.\cite{haeusler96}

The remaining part in Eq.~(\ref{slow}) can now be employed to 
provide  a microscopic definition of the {\em capacitance} $C$ of a
strong impurity in a quantum wire.  Since $f(x\to \pm \infty)=0$,
we have with $Q= -e \int_0^\infty dx [\rho(x)-\rho(-x)]/2$
the simple result $C_{\rm fr}=Q/U_{\rm ex}=ef(0)/\sqrt{\pi} U_{\rm ex}$,
or explicitly
\begin{equation}\label{capacitance}
C_{\rm fr}= (e/\pi )^2 \int \frac{dk}{2\pi} \frac{U_0-U_k}{\omega_k^2}\;.
\end{equation} 
There is also a  contribution
$C_{2k_F}$ due to the Friedel oscillation, i.e., the 
$2k_F$ component of $\rho$ in Eq.~(\ref{dens}).
While the finite-range capacitance
$C_{\rm fr}$ vanishes for the usual zero-range Luttinger liquid,
$C_{2k_F}$ is finite even for $R=0$ unless one is in the
non-interacting limit $g=1$. 
The result for $C_{2 k_F}$ can be found in Ref.\onlinecite{egger96}.
In total, since the charges simply  add up,
the capacitance is then given by $C=C_{\rm fr} + C_{2k_F}$. 
For the rather long-ranged interactions typically present in quantum wires,
 $k_F R \gg 1$, the slow component will dominate, $C_{\rm fr} \gg  C_{2k_F}$.

To give a concrete example, we consider the particularly simple
form of an exponential interaction with dimensionless 
forward scattering strength $u=U_0/\pi v_F$ such that
$g=1/\sqrt{1+u}$,
\begin{equation}\label{spec}
U(x-y) = \frac{\pi v_F u}{2R} \exp(-|x-y|/R) \;,\quad 
U_k=\frac{\pi v_F u}{1+(Rk)^2} \;,
\end{equation}
which allows for explicit analytical calculations. 
The full density profile is found
from Eqs.~(\ref{fk}) and (\ref{slow}),
\[
\rho_0(x)= -\frac{eU_{\rm ex}}{2\pi v_F} {\rm sgn}(x) [
1+  u e^{-|x|/gR } ]\;,
\]
such that the impurity charge at $x=0$
is exponentially screened on a scale $gR$. 
The finite-range capacitance
is easily found from Eq.~(\ref{capacitance}) as 
\begin{equation} \label{c0}
C_{\rm fr} = \frac{e^2 u R }{2\pi v_F \sqrt{1+u}} \;.
\end{equation}
Interestingly, the capacitance is proportional to the 
interaction range, $C_{\rm fr}\sim R$. This behavior follows immediately from  
simple dimensional scaling if the Coulomb interaction depends only
on $|x-y|/R$.
Furthermore, $C_{\rm fr}$ increases monotonously
(but not linearly) with Coulomb interaction strength $u$.

From Eqs.~(\ref{wk}) and (\ref{bos}), one
 can also evaluate the boson propagators in closed form for 
the interaction (\ref{spec}).  The lengthy result can be simplified in
two important limits. For $|\omega|\ll v_F/R$, the Luttinger
liquid result
\begin{equation}\label{apprb}
F(x,\omega) = \frac{g\pi}{|\omega|} \exp[-|g\omega x/v_F|] 
\end{equation}
is recovered. On the other hand, for $|\omega| \gg v_F/R$,
one finds the non-interacting result, i.e., Eq.~(\ref{apprb}) with
$g=1$. Apparently there is a new energy scale $v_F/R$ associated
with the interaction range. For frequencies
small compared to this scale, the electrons basically see a zero-range
(Luttinger liquid) interaction, while for frequencies larger than $v_F/R$,
the non-interacting behavior is found.  In that case the electrons
are too fast to see each other via the finite-range Coulomb interaction.

The Friedel oscillation can be evaluated from Eq.~(\ref{envelope}),
since $W(x)$ directly determines the $U_{\rm ex}=0$ Friedel oscillation for a 
strong scatterer,\cite{foot2}
\[
\langle \rho_{2k_F}(x) \rangle= - (k_F/\pi) W(x) \sin[2k_F x] \;.
\]
Splitting up the frequency integration in Eq.~(\ref{envelope})
into $|\omega| < v_F/R$ and
$|\omega| > v_F / R$, and using the respective
boson propagators, one finds that approximately
\[
W(x)\simeq  (1+2gx/R)^{-g} \, \frac{2k_F x}{1+2x/R} \;.
\]
For $x\gg R$, this reproduces the Luttinger liquid result,
namely an algebraic decay of 
the Friedel oscillation $\sim x^{-g}$.
On the other hand, there is a crossover to the faster non-interacting
law for $x\ll R$, where the Friedel oscillation decays as $1/x$.
We note that for a weak impurity, the situation is more complex
since there is a competing influence trying to slow down the 
Friedel oscillation close to the impurity.\cite{egger95}

A similar behavior characterizes the conductance. The Luttinger
liquid power laws \cite{kane92,matveev93,fendley95} are
observed only on energy scales small compared to $v_F/R$,
while one has the non-interacting behavior for larger energy
scales.  

Finally, let us comment on the case of an unscreened interaction
of the form  \cite{longrange}
\begin{equation}\label{lr}
U(x-y)=  \frac{e^2}{\kappa \sqrt{(x-y)^2+d^2}} \;,
\quad U_k = -\frac{2e^2}{\kappa} \ln[kd] \;.
\end{equation}
Here, $\kappa$ is the dielectric constant and $d$ denotes the
width of the wire ($kd\ll 1$).  The capacitance can be obtained from 
Eq.~(\ref{capacitance}). Since the interactions are strong,
Eq.~(\ref{capacitance}) can be simplified to
\[
C_{\rm fr} = - \frac{e^2}{\pi^2 v_F} \int_{1/R}^{1/d} \frac{dk}{k^2}
\left( 1+ \frac{\ln[R/d]}{\ln[kd]}\right) \;,
\]
where a finite interaction range $R$ allows for a 
controlled evaluation of $C_{\rm fr}$.  The integration yields 
\begin{equation} \label{clr}
C_{\rm fr}
 = \frac{e^2}{\pi^2 v_F} \frac{R}{\ln [R/d]} + {\cal O}(R/\ln^2[R/d]) \;.
\end{equation}
Therefore $C_{\rm fr}$ diverges $\sim R/\ln R$ as $R\to \infty$
for an unscreened Coulomb interaction. This underlines the crucial importance 
of charging effects in quantum wires with a long-ranged $1/|x-y|$
Coulomb potential.\cite{maurey96} It is noteworthy that due 
to the $R/\ln R$ dependence of the capacitance (\ref{clr}), the 
charge $Q_s$ disappears in the unscreened case: 
The interaction range $R$ becomes larger than the
length of the quantum wire only in the absence of screening gates.

\section{Double barrier}  \label{sec:4}

The capacitance $C$ and hence Coulomb blockade effects
can be experimentally observed once islands
are present in the quantum wire.  The simplest case is given by
a double barrier.\cite{kane92,furusaki93,sassetti95}
We consider two strong impurities
located at $x=\pm d/2$, take
an infinitesimal two-terminal voltage $U_{\rm ex}$ and compute the 
resonance condition as a function of the gate voltage $\varphi_G$ 
coupling to the island charge.
Naturally, such a set-up might be difficult to realize experimentally.
Nevertheless, the calculations for this model show how one can 
observe charging effects in
principle. The great merit of such a set-up is that thermodynamic
calculations  suffice to determine the location of the 
resonances.\cite{vanHouten92}
 Since the large barriers confine the charge on the island
 to some integer value $n$, the resonance condition at $T=0$
is simply $E(n) = E(n-1)$, where $E(n)$ is the energy of the total 
system with charge $ne$ on the island.

We start from $H=H_0+H_I + H_S+H_G$, where $H_S$ describes the impurities
and $H_G$ the coupling to the gate voltage $\varphi_G$,
\[
H_G= e\varphi_G \, [\theta(d/2)-\theta(-d/2)]/\sqrt{\pi}\;.
\]
Since the gate voltage couples to the island charge
capacitively, there is no need to resort to a boundary condition and
one can use the standard term $H_G$.
Furthermore, we assume that the impurities 
are strong enough to pin the plasmon displacement fields to the minima
of the cosine, such that the fields $Q(\tau)$ and $N(\tau)$ defined by
\begin{eqnarray*}
Q &=& [\theta(d/2)+\theta(-d/2)]/\sqrt{\pi} \\
N&=& [\theta(d/2)-\theta(-d/2)]/\sqrt{\pi}  
\end{eqnarray*}
become discrete. Neglecting the Friedel oscillation, the charge on
the island is $n=k_F d+ N$, while $Q$ is associated with transport
through the island. 

Enforcing the definitions of $Q$ and $N$ by Lagrange
multiplier fields, one can proceed as before.
Inserting the solution of the Euler-Lagrange equation into the 
action and integrating out the Lagrange multipliers yields the
effective action\cite{furusaki93}
\begin{eqnarray*}
S  &=& \frac{\pi^2}{2} \int \frac{d\omega}{2\pi} \left(
\frac{N(\omega)N(-\omega)}{F(0,\omega)-F(d,\omega)}
+\frac{Q(\omega)Q(-\omega)}{F(0,\omega)+F(d,\omega)}
\right) \\ && \qquad + e\varphi_G \int d\tau\, N(\tau) \;.
\end{eqnarray*}
The mode $N$ is gapped while $Q$ is ungapped
and hence irrelevant with respect to charging effects.
The low-frequency sector can therefore be described by an
effective energy 
\[
E(N) = N^2/2A + e\varphi_G N \;,
\]
where we have introduced the quantity $A=[F(0,0)-F(d,0)]/\pi^2$
whose explicit form is 
\begin{eqnarray}\label{defa}
 A & = & (v_F/\pi^2) \int dk \, [1-\cos(kd)]/\omega_k^2 \\
&=& \frac{g^2 d}{\pi v_F} + \frac{g^2}{\pi^3} \nonumber
\int dk \,[U_0-U_k]\,\frac{1-\cos[kd]}{\omega_k^2}\;.
\end{eqnarray}
The first part is the renormalized level spacing derived
in Refs.\onlinecite{kane92,furusaki93}, while the second
part is an additional contribution due to the charging energy.
The  condition $E(N) = E(N-1)$ then directly yields the
spacing of subsequent values of $\varphi_G$ where one has  a resonance,
\begin{equation}\label{spacing}
e \Delta \varphi_G =  1/A\;.
\end{equation}

For interaction range $R\ll d$, the $\cos[kd]$ in Eq.~(\ref{defa}) does
not contribute, and one recovers the single-impurity capacitance $C_{\rm fr}$
defined in Eq.~(\ref{capacitance}). Taking into account also the
Friedel oscillation contribution $C_{2 k_F}$,
one has the resonance condition in a particularly simple form,
\begin{equation} \label{sp2}
e g^2\Delta \varphi_G=
\left[\frac{ d}{\pi v_F} + \frac{2 C}{e^2}\right]^{-1}\;.
\end{equation}
Remarkably, the capacitance will decrease the spacing $\Delta \varphi_G$,
and therefore $\Delta \varphi_G$ is {\em smaller}\, 
 than predicted for the zero-range model.\cite{kane92,furusaki93}
At first sight, this might appear counter-intuitive because charging effects
supposedly increase the spacing.\cite{vanHouten92}
However, since $A$ is always diminuished by repulsive 
interactions, see Eq.~(\ref{defa}),
interactions {\em per se}\, will always enhance the spacing $\Delta \varphi_G$.
The factor $g^2$ in Eq.~(\ref{sp2}) tends to increase $\Delta\varphi_G$,
 while  a
finite capacitance $C$ decreases the spacing again. 
In total, however, one is still left with an
enhanced spacing compared to the resonant-tunneling value
$e\Delta \varphi_G = \pi v_F/d$ of a non-interacting wire.
As a simple example, we consider Eq.~(\ref{sp2}) for the
exponential interaction potential (\ref{spec}) leading to
the capacitance (\ref{c0}). In that case, Eq.~(\ref{sp2}) becomes
with the dimensionless interaction strength $u=U_0/\pi v_F$
\[
e\Delta \varphi_G = \frac{\pi v_F (1+u) }{d + R u/\sqrt{1+u}} \;.
\]
For a given interaction range $R$, increasing the
interaction strength $u$ will always lead to a larger spacing
$e \Delta \varphi_G$ compared to the value
$\pi v_F/d$ arising from the bare level spacing of the island. 
On the other hand, for given $u$, increasing $R$ decreases $\Delta \varphi_G$
for the reasons discussed above.

\section{Conclusions} \label{sec:5}

In this paper, we have employed the boundary condition formalism
to investigate charging effects in one-dimensional quantum wires. 
Here, charging effects have two distinct origins:
(i) The finite range of the screened Coulomb interaction potential,
and (ii) the long-ranged nature of the Friedel oscillation in
a Luttinger liquid. The full density profile around an impurity in the
presence of external voltage sources can be computed using
the bosonization method under appropiate boundary conditions.
Using that result, one can infer the value of the capacitance of the
impurity. While this capacitance and hence charging effects
do not seem to have consequences for dc transport properties
 in the case of a single impurity, the
condition for resonant tunneling through a double barrier is modified
compared to previous estimates if charging is properly taken into account. 

\acknowledgements

We wish to thank W. H\"ausler and F. Kassubek for useful discussions.


\begin{references}
\bibitem{foot1} This amounts to $\int d\omega \langle q(\omega)\rangle
\, \partial_x F(x,\omega)/F(0,\omega)  =0$
which can be checked by noting that $\langle q(\omega\neq 0) \rangle=0$
and $\partial_x F(x,0)/F(0, 0 ) = 0$.
\bibitem{foot2} The pinning function of Ref.\onlinecite{egger95} is 
equal to unity for a strong scatterer. This holds for
any repulsive Coulomb interaction potential.
\end{references}
\end{document}